# Ultrafast Green Single Photon Emission from an InGaN Quantum Dot-in-a-GaN Nanowire at Room Temperature

*Swagata Bhunia[1,2], Ayan Majumder[2], Soumyadip Chatterjee[2], Ritam Sarkar[2], Dhiman Nag[2], Kasturi Saha[2,3], Suddhasatta Mahapatra[1,3,\*], Apurba Laha[2,3,\*]*

1. Department of Physics, Indian Institute of Technology Bombay, Powai, Mumbai-400076

2. Department of Electrical Engineering, Indian Institute of Technology Bombay, Powai, Mumbai-400076

3. Centre of Excellence in Quantum Information Computing Science & Technology, QuICST.

ABSTRACT: In recent years, there has been a growing demand for room-temperature visible single-photon emission from InGaN nanowire-quantum-dots (NWQDs) due to its potential in developing quantum computing, sensing, and communication technologies. Despite various approaches explored for growing InGaN quantum dots on top of nanowires (NWs), achieving the emission of a single photon at room temperature with sensible efficiency remains a challenge. This challenge is primarily attributed to difficulties in accomplishing the radial confinement limit and the inherent giant built-in potential of the NWQD. In this report, we have employed a novel Plasma Assisted Molecular Beam Epitaxy (PAMBE) growth approach to reduce the diameter of the QD to the excitonic Bohr radius of InGaN, thereby achieving strong lateral confinement. Additionally, we have successfully suppressed the strong built-in potential by reducing the QD diameter. Toward the end of the report, we have demonstrated single-photon emission ($\lambda$=561 nm) at room-temperature from the NWQD and measured the second-order correlation function $g^2(0)$ as 0.11, which is notably low compared to other reported findings. Furthermore, the lifetime of carriers in the QD is determined to be 775 ps, inferring a high operational speed of the devices.

Introduction:

A room-temperature single-photon source operating in the visible wavelength is highly in demand for the development of quantum technologies in computing, sensing, and communication.[1-6] Further, the necessity for a single-photon in the visible wavelength range arises from the availability of high-speed Si single-photon detector and existing fiber and polymer materials with a transmission window in the visible wavelength range. Among various potential candidates, the III-Nitride nanowire-quantum-dots (NWQDs), particularly the InGaN NWQD, stands out as a promising choice due to its stability,[7,8] wavelength tunability,[9-11] higher band offset with other III-Nitride material, fast generation rates[12] and the ability to emit single photons in response to applied electrical signals.[13] Furthermore, due to the valence band mixing effect, the InGaN NWQDs emit linearly polarized single photon (polarization perpendicular to c-axis), which is highly desirable for quantum key distribution and linear optical quantum computing.[14-16] Additionally, the NWs geometry provides many advantages, including dislocation-free structure for quantum dot growth,[17] higher light extraction efficiency,[18] and a lower polarization field when compared to quantum wells. [19] This reduced polarization field is due to the structural relaxation of QDs along the sidewalls of the NWs.[20, 21] Moreover, the reduction of polarization field in the NWQDs allows higher In incorporation in the InGaN QD, enabling access to the visible to IR-wavelength regime.[11, 22] Another advantage is that the NWQD need not have a wetting layer underneath the InGaN/GaN QDs, which would otherwise hinder the high temperature-operation of SPS, as carriers may escape through it. [21, 23]

Significant progress has been made in the field of InGaN NWQD-based single photon emitting devices. For example, single photon emission has been achieved from an electrically driven InGaN NWQD, [13] optically excited QD in nanopillar, [24, 25] sidewall controlled InGaN QD, [7] and even single photon emission from non-polar and semi-polar InGaN QD. [12, 26] The highest temperature, at which the single photon emission has so far been reliably detected reached 220 K. [9] So, the emission of single photon at room temperature from these InGaN NWQD devices still remain unaccomplished. The primary challenges in achieving room-temperature single photon emission from these NWQDs are the limited control to the radial quantum confinement regime and the presence of a giant built-in potential. To enable room temperature operation, it is essential to have three-dimensional (3D) confinement within the QDs, and for that purpose, QDs with extremely small dimensions are necessary. While reducing the vertical dimension of NWQD below the excitonic Bohr radius (3 nm for GaN and 8 nm for InN [27]) poses minimal challenges, attaining lateral dimensions down to the excitonic Bohr radius remains a formidable task, rendering lateral confinement inaccessible. Additionally, when QDs are grown along the (0001) direction (c-plane), the strong built-in potential emerges, which in turn weakens the strength of the exciton oscillator. [28] This reduction in strength results in a reduced probability of radiative transition, creating uncertainty of single photon emission at room temperature from these NWQD devices. [29] Additionally, the reduction of strength of the exciton oscillator leads to an extension of carriers' lifetime, which causes a significant reduction in device operating speed. [26]

In this report, using a unique growth technique, we have grown the QDs on top of the nanowire-supported quantum wires and narrowed down the diameter of QD to the sub-10-nm limit. Thus, by accessing the radial confinement-limit, we have demonstrated room-temperature single photon emission from these InGaN NWQDs in the visible wavelength (~ 561 nm) regime. Furthermore, by suppressing the built-in potential of the QD through diameter reduction, we have reduced the carriers' lifetime to the order of picosecond (~ 775 ps).

Experimental:

All NWQDs and quantum disk samples were grown on sapphire substrate using the RIBER C21-Plasma Assisted Molecular Beam Epitaxy tool. The details of NW growth, and on top of that the quantum wire growth, can be found in our previous report. [30] By combining a controlled thermal decomposition, and its inhibition by an AlN capping layer, we have grown the GaN quantum wires on top of the GaN NWs. Following the unconventional approach, the tip diameters of quantum wires were reduced to a sub-10-nm limit, and on top of these quantum wires the InGaN QDs were grown. While the NWs were grown at 880 °C (for 2 hrs 30 min), the InGaN QD was grown at 550 °C. The InGaN QDs were grown for 3 minutes followed by immediate encapsulation with GaN, which was also grown for 3 minutes. Since InGaN dissociates at higher temperatures, in comparison to the growth temperature of GaN, the QD were grown at a lower temperature. An alternative set of samples was prepared for comparative analysis, in which the InGaN was directly grown on GaN NWs without preceding any decomposition process. As in the former method, the InGaN was capped with a low-temperature GaN layer; this later configuration is referred to as InGaN quantum disk. In all cases, during the growth of InGaN, the In and Ga fluxes were maintained at $1.7 \times 10^{-7}$ Torr and $9.2 \times 10^{-8}$ Torr, respectively, while the $N_2$ flow of the plasma cell was kept constant at 2.0 sccm. However, during the growth of GaN NWs, although the $N_2$ flow rate remained the same, the Ga flux was maintained at $2.1 \times 10^{-7}$ Torr, which is higher than the Ga flux of QD or quantum disk growth. This is due to the fact that the NWs were grown at a higher temperature, where the Ga desorption rate is considerably high.

The Microstructural properties of the NWQD samples and quantum disks samples were characterized by Field Emission Scanning Electron Microscopy (FESEM) and High Resolution Tunneling Electron Microscopy (HRTEM). The optical properties of both sets of samples were investigated by temperature-dependent Photoluminescence Luminescence (PL) measurement, where a laser source with a wavelength of 325 nm and an 80 um spot size was used. For single photon characterization, the second order correlation function was measured by Hanbury Brown Twiss (HBT) experiment.

Results and Discussions:

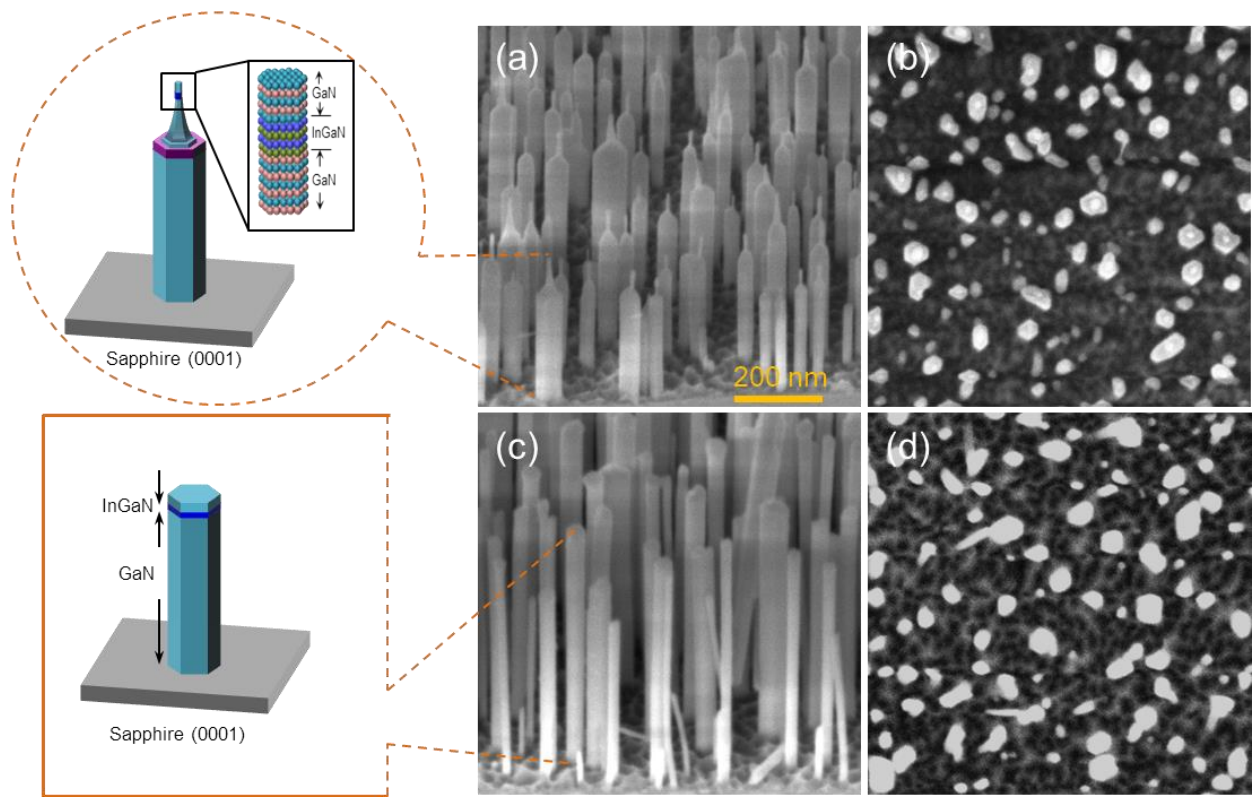

**Figure 1.** While (a, c) show the SEM 45°-tilted view images of the NWQD and nanowire-quantum-disk samples (b, d) exhibit the SEM-top view images of the same samples. The quantum wires on top of NWs are observed for NWQD sample (a), however, the quantum wires are absent in the case of the nanowire-quantum-disk sample (b). Although the existence of tiny dots that correspond to quantum wire are noticeable in (c), they are not observed in (d).

The SEM 45°-tilted view image shows the quantum wire on top of the NW (fig. 1(a)). The average lengths of the NWs and quantum wires are 391.82 nm and 82.51 nm, respectively. It can be seen that on top of each NW, the quantum wires are grown, and a significant reduction in the diameter of the quantum wires is clearly noticeable. From the SEM top-view image, the existence of quantum wire (tiny dot) on top of NW and the difference in their diameters can also be noticed (fig. 1(b)). The average diameter of NWs and quantum wires are measured as 54.12 nm and 15.13 nm, respectively. However, the actual diameter at the top portion of quantum wire cannot be measured by SEM as it is beyond the resolution of our SEM.

On the other hand, the SEM-tilted-view (fig. 1(c)) and top-view images (fig. 1(d)) of nanowire-quantum-disk sample show the NWs of length of 534.20 nm and diameter of 52.89 nm, respectively. No quantum wires were observed on top of the NWs. Moreover, the diameters of the NWs remain well above the excitonic Bohr radius of InGaN. The details of the NWQD structure and the NW-quantum disk structure are depicted inside the dotted circle and the square, respectively. In the NWQD image, on top of the quantum wire, the InGaN QD embedded in GaN is delineated.

The HRTEM image of a single NWQD reveals the quantum wire positioned atop the NW (fig. 2(a)). This image clearly illustrates a noticeable difference in diameters between quantum wire and NW. The average diameters of the quantum wire and NW are measured as 50.63 nm, and 9.58 nm, respectively. Notably this observation indicates a significant five-time reduction in the quantum wire's diameter, which has been achieved through the implementation of our unique growth approach. Furthermore, a higher-resolution image acquired from the tip region of the NWQD presents the InGaN QD with diameters of 9.58 nm and heights of 2.88 nm (fig. 2(b)). So, the diameter of the QD is below the excitonic Bohr radius of the InGaN, this achievement was hitherto unattainable due to the inherent complexity of the growth process. Significantly, the QDs possess no threading dislocations, no observable dislocation lines extending from the NWs to the QD (fig. 2(c)). Even though the GaN capping layer was grown at lower temperatures, no dislocation lines are visible. This implies that surroundings of the QD are free from any defects. Additionally, no InGaN wetting layer is observed beneath the QD. Therefore, one would anticipate a reduction of spectral diffusion within the QD, which could enhance the quantum efficiency of NWQD devices and facilitate single photon emission.

On the other hand, HRTEM of single NW quantum disk is shown in fig. 3(d); no quantum wires are observed on top of the NW. The higher resolution image at the top portion shows the quantum disk of diameter 49.75 nm, and height 2.76 nm. Moreover, the higher magnification image of the region marked with red color shows the InGaN QD, which is embedded in GaN.

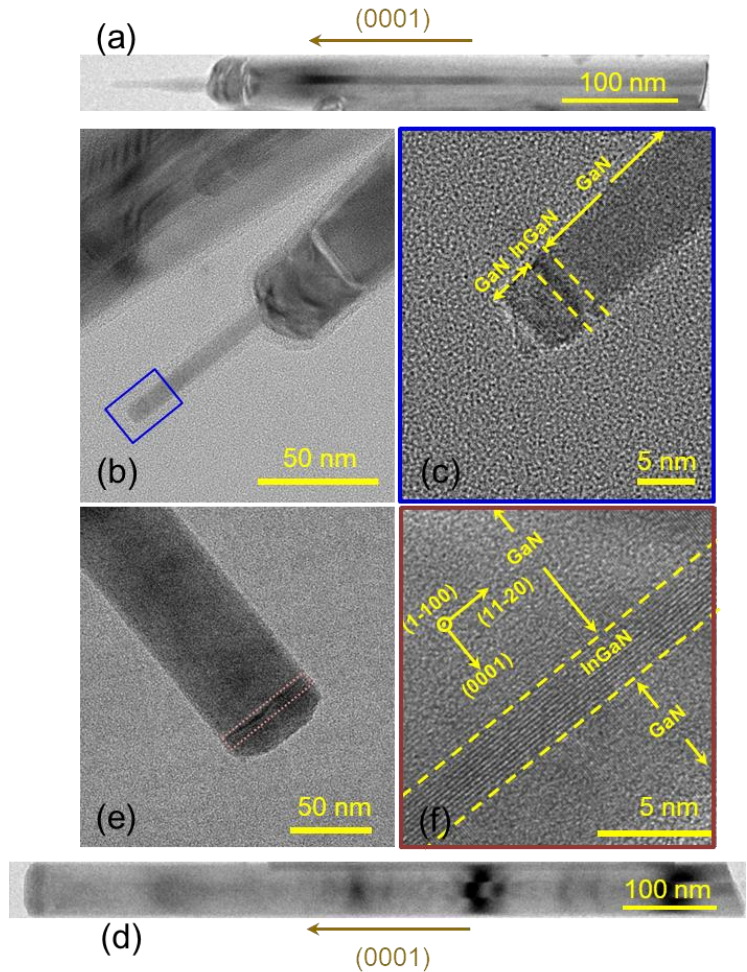

**Figure 2.** While (a), (b), (c) depict TEM images of an entire NWQD, the top portion of NWQD, and the tip of quantum wire, (d), (e), (f) presents images of a complete nanowire quantum disk, the upper segment of the nanowire quantum disk, and a higher magnification view of the quantum disk. The higher resolution image at the quantum wire tip reveals the QD with a diameter 9.58 nm, and a height of 2.88 nm (c), while the tip of quantum disk displays a quantum disk with a diameter 49.75 nm, and height 2.76 nm (e, f).

The PL spectra of InGaN QDs and quantum disks ensemble were investigated with temperature, and the results are shown in fig. 3(a,b). At 10 K, the excitonic emission peak of InGaN QDs was measured at 561 nm and the emission peak exhibited a red shift as the temperature increased towards room temperature (fig. 3(a)). The temperature-induced variation of PL emission peak with the temperature shows an S-shaped behavior, as illustrated in the inset of fig. 3(a). This S-shaped trend is a distinct characteristic associated with the combined emission of localized excitons within the QD. [31, 32] Conversely, at 10 K, the excitonic emission peak of InGaN quantum disk was observed at 623 nm (fig. 3(b)). This indicates that the emission peak of InGaN QD is blue-shifted. Since the height of the InGaN QD remains the same as the quantum disks, the shift can be attributed to the radial confinement of the QDs. Notably, the quantum disk samples also exhibited an S-shaped variation in the PL peak with increasing temperature, as demonstrated in the inset of fig. 3(b). It is noteworthy that the S-shaped variation of QD sample and quantum disks samples are distinct in their characteristics. To quantitatively analyze the energy-variation, the experimental data was fitted to the localized-state-ensemble (LSE) model, a theoretical framework developed by Li and Xu et.al. [33-35] According to the LSE model, the PL emission energy is governed by the following equation as a function of temperature

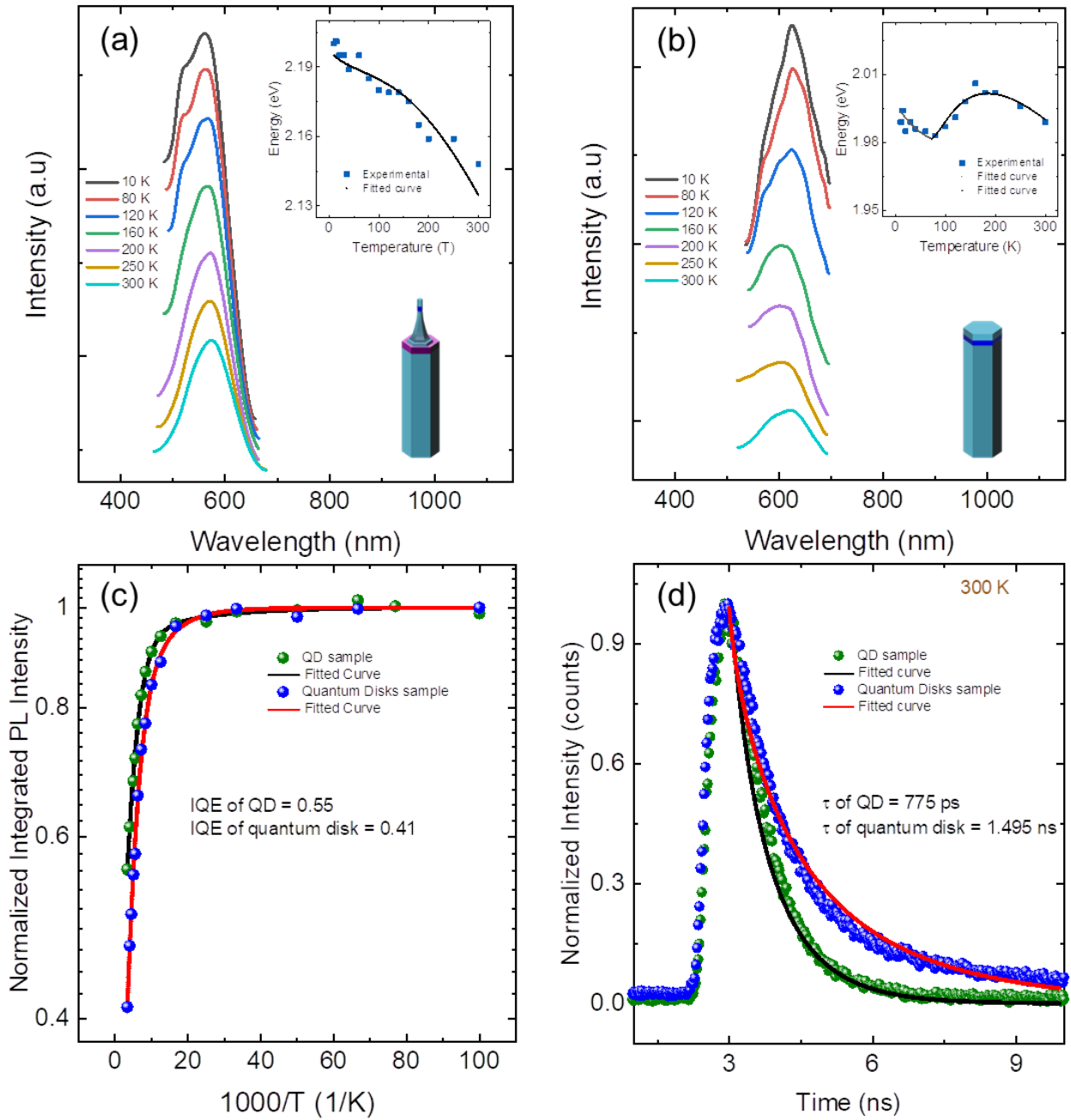

**Figure 3.** (a) and (b) display the temperature-dependent PL emission of NWQDs and quantum disks, respectively. Insets of figures illustrate the variation of excitonic emission energy with temperature for QDs (a), and quantum disks (b). (c) presents the plot of normalized integrated PL intensities of these emitters as a function of temperature, while (d) displays the decay time of carriers within the dot and the disk.

$$E(T) = E_0 - \frac{\alpha T^2}{(\beta + T)} - \gamma(T) k_B T$$

Where $E_0$, α and β represent the band gap of the quantum disk or QD at 0 K, the Varshni's parameter, and the Debye temperature, respectively. In the latter part of the equation, $\gamma(T)$ and $k_B$ signify the temperature-dependent dimensionless parameter and the Boltzmann constant, respectively. The parameter $\gamma(T)$ is determined through a numerical solution of the following equation:

$$\left\{ \frac{1}{\gamma} \left( \frac{\sigma}{k_B T} \right)^2 - 1 \right\} e^{-\gamma} = \left( \frac{\eta}{1-\eta} \right) e^{\frac{-\Delta E}{k_B T}}$$

Here, σ represents the width of localized states' distribution, which is assumed to follow a Gaussian type distribution. The terms $\eta$ and $\Delta E$ correspond to the luminescence quantum efficiency and energy difference between the exciton's occupied level and the central energy of localized states, respectively. The values of the parameters are obtained through experimental curve fitting and are presented in Table 1.

| Sample | Localization Centers | $\alpha\ (\frac{meV}{K})$ | $E_0(eV)$ | $\Delta E(meV)$ | $\sigma(meV)$ | $\eta$ |
|---|---|---|---|---|---|---|
| Quantum Disks | red | 4.71x10-4 | 1.995 | -5.9 | 11.10 | 0.083 |
|  | blue | 4.47x10-4 | 2.053 | 70.3 | 23.20 | 0.265 |
| QDs | blue | 7.95x10-4 | 2.226 | 28.3 | 23.70 | 0.47 |

Since the InGaN quantum disks and QD do not have any dislocations when grown on NWs, the emergence of localization centers is primarily attributed to the inhomogeneity in the indium content. The data presented in Table-1 reveals that the quantum disk samples exhibit two localization centers at energy levels 1.995 eV and 2.053 eV. In contrast, the QD samples display a single localization at 2.226 eV. This observation indicates a reduced compositional inhomogeneity in QDs relative to quantum disks. Consequently, the temperature-dependent behavior of the PL peak demonstrates a less pronounced 'S-shaped' variation in the QD sample. Despite having fewer localization states the efficiency ($\eta$) of QDs (0.47) is higher than that of the quantum disks (0.26), owing to the radial confinement-effect. This is further confirmed by the temperature-dependent integrated PL intensity plot, in which we can observe that compared to InGaN quantum disks sample, the decay of integrated PL intensity with temperature is less in the case of the QD sample (fig. 3(c)). The internal quantum efficiency (IQE) of InGaN QD (0.55), determined by the ratio of PL intensity at 300 K and 10 K ($\frac{I_{300\,K}}{I_{10\,K}}$)), exceeds the IQE of InGaN quantum disks (0.41). This temperature-dependent PL analysis provides direct evidence of the radial confinement regime of the InGaN QD that has been accessed by solely reducing the diameter of the QDs. Our previous study has established that reducing diameter not only facilitates access to the radial confinement regime but also mitigates the built-in potential of the QDs. The reduction in built-in potential enhances the exciton oscillator strength, resulting in a shorter carrier lifetime. The time-resolved measurements of both the samples indicate that the QD exhibits notably a faster decay time compared to the quantum disk (fig. 3(d)). For quantitative analysis, we fit the experimental curve using a stretch exponential equation: $I = I_0 \exp(-\frac{t}{\tau})^\rho$, where $\tau$ and $\rho$ denote the carrier-lifetime and stretched parameters. In the case of QD, the carrier-lifetime is found to be 775 ps, which is shorter than that of the quantum disks (1.495 ns). Even though the QD is grown along the c-direction, the carriers' lifetime is much shorter than the previously c-plane-grown QD, which is typically order of 2 to10 ns. [36, 37] This implies that we have suppressed the built-in potential in the c-plane NWQD. These results signify that the grown NWQD holds significant potential for single photon emission at room temperature. Additionally, the shorter carrier lifetime suggests that the device can achieve a faster repetition of single photon emission.

We carried out the single photon measurement using the PicoQuant Microtime 200 set-up, as depicted schematically in fig. 4(e). Firstly, the NWQDs were transferred onto a Si (111) substrate and then excited with a pulsed laser operating at a wavelength of 405 nm, with a pulse rate of 40 MHz, power 200 nW, and a spot size of ~ 0.3 um. The signal emitted from the QD, was splitted using a 50:50 beam splitter and collected through two distinct avalanche photo diodes (SPAD). The acquired signal was then fed to the Time Correlated Single Photon Counting (TCSPC) electronics, which measures the precise arrival times of photons through the Time Tagged Time-Resolved (TTTR) mode. This mode also enables the recording of the spatial origin of photons, utilizing the Transistor-to-Transistor Logic (TTL) mode, which is received from the piezo-scanner and piezo controller. Therefore, using the same set up and fluorescence lifetime imaging process, a single NWQD was located, as shown in fig. 4(a). Repeating the same process for the quantum disk sample, a single quantum nanowire-quantum-disk was also found out (fig. 4(b)).

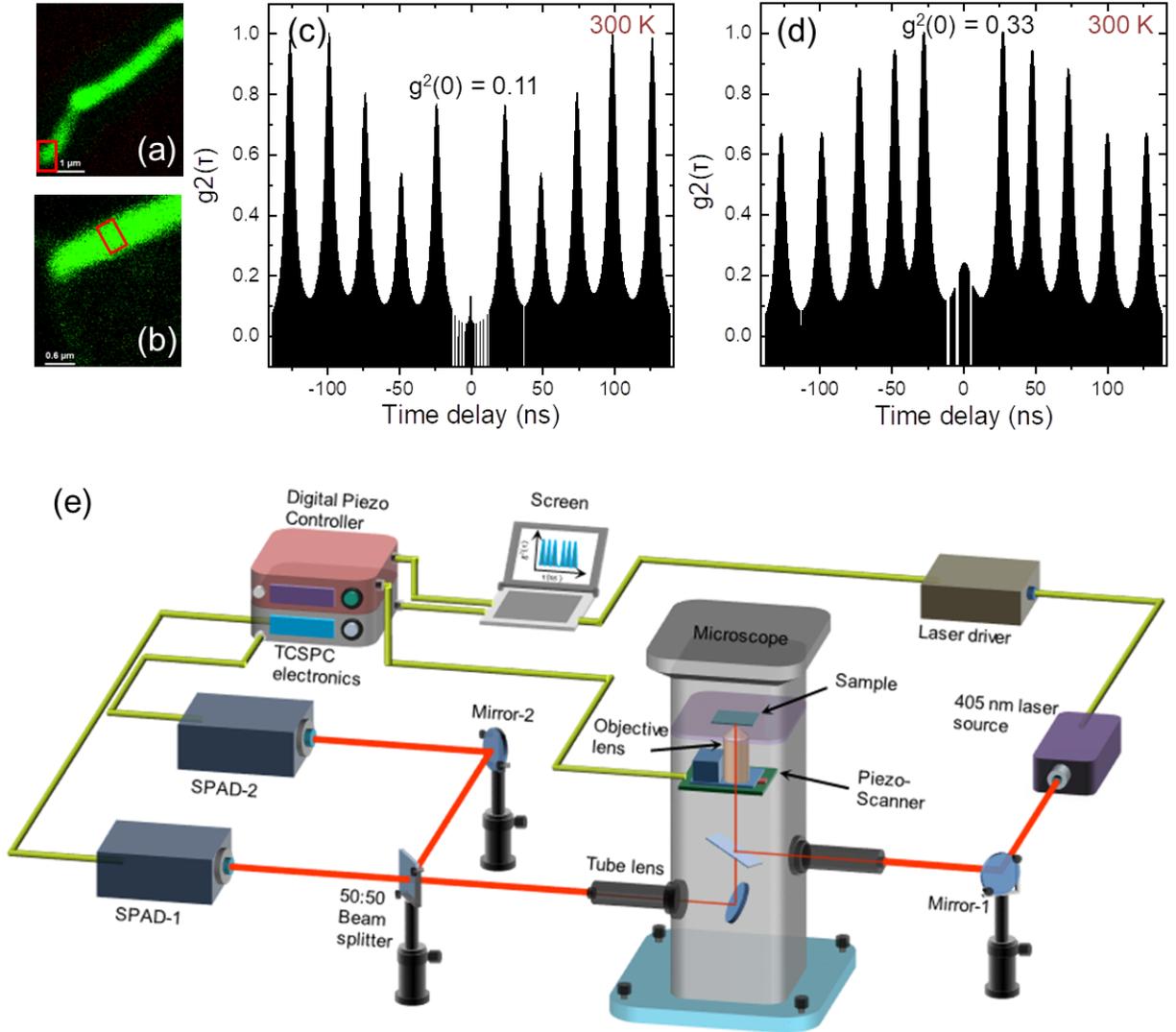

**Figure 4.** (a) and (b) show the lifetime imaging of NWQDs and quantum disks, respectively. (c) and (d) present the room-temperature second-order-correlation measurement at the selected portion of QD (fig. (a)), and quantum disk (fig. (b)), respectively. (e) displays a schematic diagram of the single-photon measurement and lifetime measurement set-up.

The single photon measurement was carried out at room temperature in the specified region of the NWQD and quantum disk. The measurement of second order correlation function $g^2(\tau)$, defined as $g^2(\tau) = \frac{<I(t).I(t+\tau)>}{<I(t)>^2}$ (where $I(t)$ is the intensity of light emitted from a QD at time $t$, and $\tau$ is the delay time), assesses the single photon emission characteristics (photon statistics). At a zero-time delay, the value of $g^2(0) < 0.5$ affirms the single photon emission. In the case of QD, the plot of $g^2(\tau)$, with time delay $\tau$, is presented in fig. 4(c). The $g^2(\tau)$ exhibits a clear antibunching at $\tau = 0$, indicating a suppression of the photon coincidence counts. Furthermore, the measured value of $g^2(0) = 0.11$, demonstrating the emission of a single photon even at room temperature. Remarkably, $g^2(0)$ remains exceptionally low without the introduction of Distributed Bragg Reflectors (DBRs) or background emission corrections, implying a high single photon emission efficiency of the NWQD. For further analysis, at room temperature, the $g^2(\tau)$ was measured at the specified region of the quantum disk (fig. 4(b)) and the measured value of $g^2(0)$ is 0.33 (fig. 4(c)). Comparatively, the value of $g^2(0)$ is almost three times less in the case of QD, indicating a significant improvement in the single photon emission efficiency due to the reduced diameter of the QD. The combined effects of extreme 3D-quantum-confinement and reduced built-in potential enhance the efficiency of the NWQD. Additionally, the absence of dislocations in the vicinity of the QD reduces the spectral diffusion measured at higher temperature. All these effects enable a single photon emission from the InGaN NWQD. Since the carrier lifetime is in the order of ps, the operating speed of the QD devices is expected to be higher in the GHz range.

Conclusion:

In conclusion, we have demonstrated visible-wavelength single-photon emission from the InGaN NWQD devices at room-temperature. The efficiency of NWQD devices is significantly higher than that of quantum disks. The enhancement in efficiency is attributed to the superior quality of InGaN-QD grown by our unique growth approach, which enables access to the strong radial confinement of the QDs and thus, suppresses its inherent built-in potential. Moreover, due to the suppression of built-in potential, the life-time of carriers in the QD is remarkably low, on the order of ps. Such exceptionally short life-time makes NWQDs as a promising candidates for future quantum computing and quantum key distributions technologies operating at GHz operating frequencies.

AUTHOR INFORMATION

**Corresponding Authors**

*suddho@phy.iitb.ac.in

* laha@ee.iitb.ac.in

ACKNOWLEDGMENT

All authors acknowledge the financial support from the Science and Engineering Research Board (SERB; project no. CRG/2018/001405) of the Department of Science and Technology (DST), and the Quantum Information Technologies with superconducting devices and Quantum Dots (Project no- RD/0120-DSTIC01-001), of the Govt. of India (GoI). The authors also thank the IIT Bombay Nanofabrication Facility (IITBNF) for technical support towards the execution of this project. The authors thank Center for Research in Nanotechnology and Science (CRNTS) for allowing us to access the Fluorescence Lifetime Imaging Microscopy (FLIM) and Single Photon Characterization instrument.